\documentclass[fleqn,usenatbib]{mnras}

\usepackage[utf8]{inputenc}

\usepackage[T1]{fontenc}
\usepackage{ae,aecompl}

\usepackage{siunitx}
\DeclareSIUnit[]\sunmass{\text{\ensuremath{M_{\odot}}}}
\usepackage{tabularx}
\usepackage{arydshln}

\usepackage{graphicx}   
\usepackage{amsmath}    
\usepackage{amssymb}    

\usepackage{pdf14}

\usepackage{etoolbox}

\newlength\figureheight
\newlength\figurewidth

\renewcommand{\d}[1]{\ensuremath{\operatorname{d}\!{#1}}}

\usepackage{soul}

\title[{Universal relations for relativistic stars}]{Universal relations
  for differentially rotating relativistic stars at the threshold to collapse}

\author[G.\ Bozzola, N.\ Stergioulas \& A. Bauswein]{
Gabriele Bozzola$^{1},$ 
Nikolaos Stergioulas$^{2}$, Andreas Bauswein$^{3}$
\\
$^{1}$Department of Physics, Università degli Studi di Milano, IT-20133 Milano, Italy\\
$^{2}$Department of Physics, Aristotle University of Thessaloniki, GR-54124 Thessaloniki, Greece \\
$^{3}$Heidelberg Institute for Theoretical Studies, D-69118 Heidelberg, Germany
}
\date{\today}

\pubyear{2017}

\begin{document}

\label{firstpage}
\pagerange{\pageref{firstpage}--\pageref{lastpage}}
\maketitle

\begin{abstract}
  A binary neutron star merger produces a rapidly and differentially rotating
  compact remnant whose lifespan heavily affects the electromagnetic and
  gravitational emissions. Its stability depends on both the equation of state
  (EOS) and the rotation law and it is usually investigated through numerical
  simulations. Nevertheless, by means of a sufficient criterion for secular
  instability, equilibrium sequences can be used as a computational inexpensive
  way to estimate the onset of dynamical instability, which, in general, is close to
  the secular one. This method works well for uniform rotation and relies on the location of turning points: stellar
  models that are stationary points in a sequence of equilibrium solutions with
  constant rest mass or angular momentum. Here, we investigate differentially rotating models (using a large number of equations of state and different rotation laws) and find that several universal relations between properly scaled gravitational mass, rest mass
  and angular momentum of the turning-point models that are valid for uniform rotation, are insensitive to   the degree of differential rotation, to high accuracy. \end{abstract}


\section{Introduction}

Neutron stars, the most dense stars known in the Universe, are unique
environments to explore physics in a regime so extreme that cannot yet be reproduced
in laboratories. Studies on those compact objects have resulted in a deeper
understanding of several areas of physics and further advances are expected with
the direct detection and analysis of the gravitational waves and electromagnetic
signals emitted by coalescing binaries. These systems, formed by two neutron
stars orbiting each other and eventually merging, are promising sources
for the second generation of gravitational-waves detectors
\citep{Accadia2011,Abadie2010,Somiya2012}. One can distinguish two different outcomes
of mergers depending on the binary mass \citep{Shibata2005,Shibata2006,Baiotti2008,Hotokezaka2011,Bauswein2013}. (1) For binary masses below some threshold mass the merger results in the
formation of a hot, massive and rapidly spinning neutron star accompanied by a
strong emission of gravitational waves. 
The remnant may eventually collapse at a later time as a result of angular momentum redistribution and energy losses by means
of mass ejection, neutrino and gravitational wave emission. This scenario is referred to as \emph{delayed collapse}\footnote{Note that within this classification the delayed collapse scenario also subsumes cases where the remnant actually remains stable.}. (2) For binary masses larger than the threshold mass the remnant is too massive to be stabilized against gravitational collapse and directly forms a black hole on a dynamical timescale. This scenario is known as \emph{prompt collapse}. The stability of the remnant against gravitational collapse (prompt versus delayed collapse) affects the gravitational, neutrino and electromagnetic emission and is thus crucial for multi-messenger observations of binary neutron star mergers \citep[see e.g.][for reviews]{Faber2012,2017PasSterg,Baiotti2017}.

In general, the stability of compact stars is determined by the mass, rotation (angular momentum and rotation law), the EOS and the
temperature. In the context of binary mergers, numerical simulations show that merger remnants initially have a complex, non-uniform velocity profile and can support masses which exceed the maximum mass of uniformly rotating stars in hydrostatic equilibrium \citep{Shibata2005b,Baiotti2008,Hotokezaka2011,Bauswein2013,Kastaun2016,Hanauske2017}. Therefore, merger remnants are often modeled and described by differentially rotating stars, which can support significantly larger masses than uniformly spinning stellar configurations \citep[see e.g.][]{Baumgarte2000, Lyford2003, Morrison2004,Kaplan2014,Rosinska2016,Bauswein2017}. Compact objects with masses exceeding the maximum mass of uniformly rotating stars are usually referred to as \emph{hypermassive neutron stars} (HMNS).

Within this work we investigate rotating stellar equilibrium models with uniform and differential rotation laws.
 A key tool for studying the stability of uniformly rotating models of relativistic stars is the
turning-point method by Friedman, Ipser and Sorkin \citep{Fridman1988} for
equilibrium sequences. A uniformly rotating equilibrium stellar model will have central energy density
$ \epsilon_c $,  gravitational mass $M$, rest mass $M_0$ and angular
momentum $J$. An \emph{equilibrium sequence} is a one-dimensional slice indexed
by some parameters of the whole space of equilibrium models. The sequences we
studied in this work are constructed by varying the central energy density
$ \epsilon_c $ and holding fixed one of the other quantities. In this case, a
\emph{turning point} in a sequence occurs when two out of three first
derivatives with respect to $\epsilon_c$ vanish. For example, in a sequence of
constant angular momentum $J$ by construction
$\partial J \slash \partial \epsilon_c = 0$, therefore a turning point is a
stationary point of the function $M = M(\epsilon_c)$. The Friedman, Ipser and
Sorkin criterion implies that for equilibrium sequences of uniformly rotating
stars at a turning point:
\begin{enumerate}
  \item the derivative of the third quantity also vanishes, \label{item:FSI-criterion-1}
  \item the sequence becomes secularly unstable.
  \label{item:FSI-criterion-2}
\end{enumerate}
The theorem is based on the assumption that the variation of the total energy
depends only on the change in rest mass and angular momentum, and not in their
redistribution (which would correspond to a second derivative). This assumption
is justified by the fact that the second effect has a longer timescale.
Moreover, the criterion holds also for hot stars with the only difference that
the triplet $(M_0,M,J)$ has to be enlarged to include entropy $S$ and thus it is
required that three first derivatives out of four vanish \cite{Kaplan2014}. For cold stars,
however, entropy is not relevant to the stability because a change in entropy
does not produce a variation in energy. In this case, point
\ref{item:FSI-criterion-1} of the theorem implies that the stationary point of
the sequences with $J$ fixed must coincide with the ones of the sequences with
$M_0$ fixed. Point~\ref{item:FSI-criterion-2}, on the other hand, implies that
the criterion is only a sufficient: it does not state that all unstable models
are located on one side of a turning point, but it guarantees that every model
on that side is unstable.
Even if point~\ref{item:FSI-criterion-2} posits that the turning-point line
marks the onset of secular instability, there are reasons to assume that also
the dynamical-instability line has to be close \citep{Takami2012, Friedman2013,
  Kaplan2014}. Indeed, as angular momentum is redistributed 
on secular timescales the star encounters the onset of
dynamical instability and collapses to a black hole. Thus, the onset of the
secular instability separates stable neutron stars from collapsing ones.

The criterion is routinely used to estimate dynamical instability for cold
neutron stars \citep[e.g.][]{Takami2012, Kaplan2014} and justifies our interest
in studying equilibrium sequences.

The Friedman, Ipser and Sorkin's criterion is analytically proved only for
uniformly rotating stars, but it has been found numerically \citep{Kaplan2014}
that even for differentially rotating ones the turning points are 
indicators of the onset of instability with some accuracy.  Despite being not as
accurate as complete numerical simulations, this method has the advantage of
being computationally less expensive, and thus a large number of stellar models
with different EOSs can be constructed and tested.

In this paper we focus our attention on the properties of differentially
rotating models for turning-point sequences (i.e. at the threshold of collapse)
of constant angular momentum or constant rest mass with varying central energy
density. We find that such models satisfy universal relations, which are (to
high accuracy) independent of the choice of the EOS and on the details of the
rotation law. These relations involve gravitational mass, rest mass and angular
momentum rescaled by a proper factor that fixes the physical scale of the system
and that is provided by the turning point for the nonrotating sequence. This extends recent work 
on uniformly rotating stars \citep{Breu2016,Lenka2017} to the case of differential rotation.

The organization of the paper is as follows. Section~\ref{sec:setup} describes
the theoretical and numerical framework used, Section~\ref{sec:results} collects
our results and presents the universal relations and our application to the
threshold mass of binary neutron star mergers. Finally, a summary is left for
Section~\ref{sec:conclusions}.
If not stated otherwise, we use geometrized units with $c = G = M_\odot = 1$,
where $c$ is the speed of light in vacuum and $G$ the gravitational constant.
\section{Setup and Numerical Method}
\label{sec:setup}

\subsection{Assumptions}
\label{sec:equil-mod}

We work in full general relativity assuming a stationary and axisymmetric
spacetime with coordinates $(x^\mu)$, metric tensor $g_{\mu\nu}$ and line
element given by
\begin{equation}
  \label{eq:spacetime_metric}
  {\d s}^2 = -\mathrm{e}^{2 \gamma + \rho} {\d t}^2 + \mathrm{e}^{\gamma - \rho}(\d \phi - \omega \d t)^2 +
  \mathrm{e}^{2\alpha}({\d r}^2 + r^2 {\d \theta}^2)\,,
\end{equation}
where $t, r, \theta, \phi$ are the \emph{quasi-isotropic coordinates} and
$\gamma,\rho,\alpha,\omega$ are four metric functions that depend on the
coordinates $r, \theta$ only \citep{Bardeen1971, Friedman2013}. An equilibrium
configuration is found by solving Einstein's field equations
\begin{equation}
  \label{eq:einstein_equation}
  \mathcal{R}_{\mu\nu} - \frac{1}{2} g_{\mu\nu} \mathcal{R}^{}_{\; } = 8 \pi T_{\mu\nu}\,,
\end{equation}
where $\mathcal{R}_{\mu\nu}$ is the Ricci tensor,
$ \mathcal{R}^{}_{\; }$ the Ricci scalar and $T_{\mu\nu}$  the
stress-energy tensor of the matter. To model the matter of a neutron star we
assume its interior to be a perfect fluid with stress-energy tensor\begin{equation}
  \label{eq:stress_energy_tensor_perfect_fluid}
  T^{\mu\nu} := (\epsilon + P)u^\mu u^\nu + P g^{\mu\nu}\,,
\end{equation}
where $\epsilon$ is the energy density, $P$ the
pressure and $u^\mu$ the four-velocity.  In addition, we consider our neutron
stars to be cold and we ignore thermal effects. This is because for the
temperature expected for merger remnants thermal effects have only a secondary effect on the
 stability of the remnant, the dominant effect being the stabilitation through differential rotation  (see, e.g. \cite{Kaplan2014}).

 We assume the star to be rotating with no meridional currents. Then, the first
 integral of the hydrostationary equilibrium equation involves the function
 $F(\Omega) := u^t u_\phi$, where $\Omega := {u^\phi} \slash {u^t}$ is the
 angular velocity inside the star, with respect to a nonrotating observer at
 infinity. The choice of this function fixes the rotation law. Here, we study
 three different rotation laws: \textit{uniform} rotation, the
 \emph{one-parameter} rotation law by \cite{Komatsu1989a, Komatsu1989b}, and the
 \emph{three-parameter} rotation law by \cite{Bauswein2017}.

\subsubsection{TOV solution}
\label{sec:tov-solution}

A special case of equilibrium sequence is the one of nonrotating models
with $J = 0$, the so-called TOV\ solutions \citep{Tolman1934, Oppenheimer1939}. We
introduce for later convenience a shorthand notation for the values of some physical
quantities of the nonrotating model with \textit{maximum mass}, which is
the turning point of this sequence:
\begin{equation}
  \label{eq:tov-turning-point}
  {M}^\star := M^{\mathrm{TOV}}_{\mathrm{max}} \quad
  {M}^\star_0 := M^{\mathrm{TOV}}_{0, \  \mathrm{max}} \quad
   {R}^\star := R^{\mathrm{TOV}}_{ \mathrm{max}}\,.
\end{equation}
We also define the maximum compactness as the ratio
${M^\star} \slash {R^\star}$. For every rotation law, the maximum-mass TOV model
is the end-point of the quasi-radial instability sequence in the limit of zero
angular
momentum.

\subsubsection{Uniform rotation law}
\label{sec:uniform-rotation-law}

The most simple non-trivial rotation law is uniform rotation, where the whole
star spins as a rigid body about its axis. This is a good approximation to study
old neutron stars. Indeed, angular momentum redistribution, magnetic braking and
shear viscosity have the effect of driving the angular velocity profile towards
uniform rotation. The assumption of uniform rotation is not accurate when
applied to nascent neutron stars or to binary neutron star merger remnants,
where differential rotation modifies their bulk properties. Uniform rotation can
increase the maximum allowed mass by up to $\sim \num{25}\%$
\citep{Friedman1987}. Equilibrium configurations with rest mass larger than
${M}^\star_0$ are called \emph{supramassive}, since they reach the threshold to
black hole collapse (rather than the TOV sequence) when spun down.

\subsubsection{One-parameter rotation law}
\label{sec:one-param-rotat}

The most investigated rotation law is the
\emph{one-parameter} law, introduced in \citet{Komatsu1989a,
  Komatsu1989b}. This differential rotation is defined by:
\begin{equation}
  \label{eq:one-parameter-law}
  F(\Omega) := A^2(\Omega_c - \Omega) = \frac{(\Omega - \omega) r^2 \sin^2 \theta \mathrm{e}^{-2\rho}}{1 - (\Omega - \omega)^2 r^2 \sin^2 \theta \mathrm{e}^{-2\rho}}\,,
\end{equation}
where $\Omega_c$ is the angular velocity on the axis of rotation. The parameter
$ A $ has the units of length: it is the length-scale over which the angular
velocity varies. Thus, for $ A \to + \infty $ uniform rotation is recovered. A
dimensionless parameter is defined as:
\begin{equation}
  \label{eq:parametrs-of-one-parameter-law}
  \hat{A} = \frac{A}{r_e},
\end{equation}
where $r_e$ is the coordinate radius at the equator. From a mathematical point
of view, the one-parameter law is well-defined for every $\hat{A} \geq 0$, but
there are physical reasons to consider additional constraints on the range of
allowed values. A very small value of $ \hat{A} $ would describe a star in
which a small region near the axis would spin very fast, with the remaining star
rotating slowly. For this reason, we consider only models with
$\hat{A} > \num{0.5}$ as being astrophysically relevant. Equilibrium models that
rotate with this rotation law can support about $\num{50}\%$ more mass
than the maximum mass nonrotating model \citep{Baumgarte2000, Lyford2003,
  Morrison2004, Rosinska2016}. The term \emph{hypermassive neutron star} is used
to refer to a star that has more mass than the maximum allowed for uniformly
rotating supramassive stars.

\subsubsection{Three-parameter rotation law}
\label{sec:three-param-rotat}

The third rotation law we consider is the \emph{three-parameter} law
introduced in \citet{Bauswein2017}. This is specified by:
\begin{equation}
  \label{eq:three-parameters-law}
  \begin{cases}
    F(\Omega) = A_1^2(\Omega_c - \Omega) + (A_2^2 - A_1^2)(1-\beta)\Omega_c, & \text{if } \Omega \leq \beta \Omega_c, \\
    F(\Omega) = A_2^2(\Omega_c - \Omega),                           & \text{if } \beta \Omega \leq \Omega \leq \Omega_c,
  \end{cases}\,
\end{equation}
which is practically a piecewise extension of the one-parameter law
(\ref{eq:one-parameter-law}) to two different regions inside the star. The
regions are separated by the value $\Omega = \beta \Omega_c$. For angular
velocities larger than this, the rotation law coincides with
(\ref{eq:one-parameter-law}), with parameter $A=A_2$, and for smaller angular velocities it coincides
with (\ref{eq:one-parameter-law}) (up to a constant), with $A=A_1$. The constant
is fixed to ensure continuity of $F(\Omega)$ at the transition
$\Omega = \beta \Omega_c$. Moreover, if $\beta=1$, or if ${A}_1 = {A}_2$, the rotation law
(\ref{eq:three-parameters-law}) reduces to the usual one-parameter law (\ref{eq:one-parameter-law}). The
three-parameter rotation law allows for an astrophysically relevant core to have
a slower rotating envelope. In this way, higher values of mass are allowed
before encountering the mass-shedding limit.
\subsubsection{Numerical setup}
\label{sec:code}

We use a recent version of the \texttt{RNS} code \citep{Stergioulas1995}, which
has been extended to include models with the one-parameter rotation law
(\ref{eq:one-parameter-law}) \citep[see][]{Stergioulas2004} and with the three-parameter
rotation law (\ref{eq:three-parameters-law}) see \citep[see][]{Bauswein2017}. A
specific equilibrium model is constructed by chosing the EOS and the rotation law
and specifying the central energy density $\epsilon_c$ and the ratio of the
polar coordinate radius to equatorial coordinate radius, $r_p/r_e$. Sequences of models with desired
properties (such as a given rest mass or angular momentum) can be constructed
through an iterative procedure. We set up \texttt{RNS} to use a grid of
\texttt{DMDIV = 251} and \texttt{DSDIV = 501} (angular times radial grid points)
and to reach an internal accuracy of $\num{e-7}$ and a tolerance in the desired
value of the parameters of $\num{e-4}$ \citep[see][for more information on \texttt{RNS}]{Nozawa1998, Friedman2013}.
\subsection{Equations of state}
\label{sec:eos}

\begin{figure}
  \centering
  \setlength\figureheight{0.85\columnwidth}
  \setlength\figurewidth{\columnwidth}
   \includegraphics{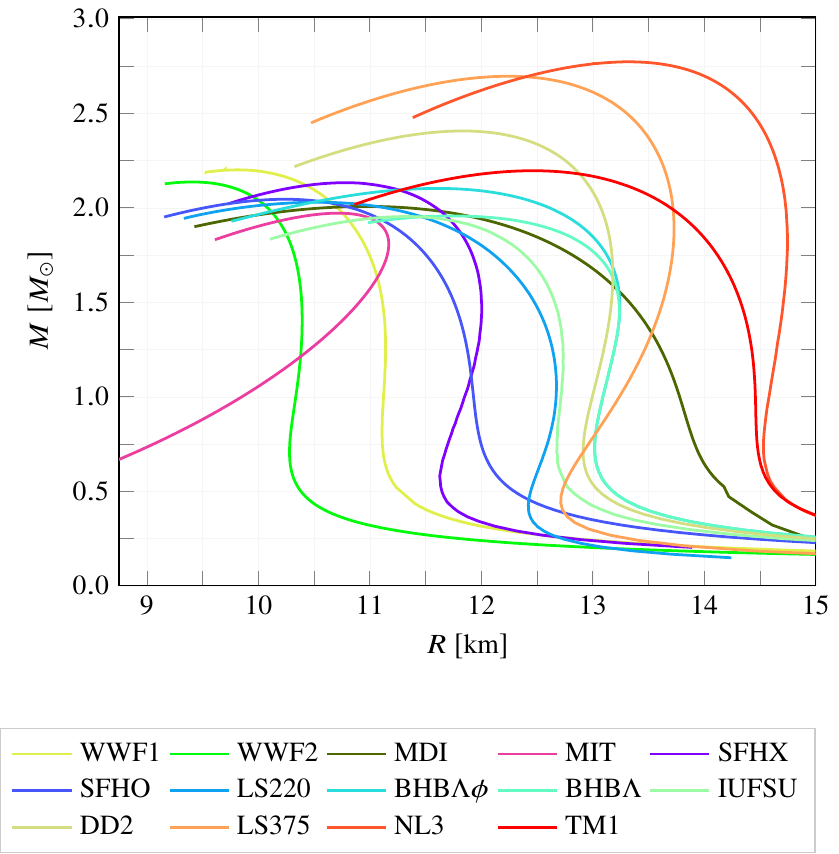}
  \caption{Gravitational mass and radius relation for non-rotating stars with
    all the tabulated EOSs shown in Table \ref{tab:eos}.
 All EOSs
    produce stellar models that are compatible with the observational mass
    limit of $\SI{2}{\sunmass}$ \citep{Demorest2010,Antoniadis2013}. }
  \label{fig:eos}
\end{figure}

We consider thirteen cold tabulated EOSs in neutrino-less $\beta$-equilibrium,
which are all compatible with the observational requirement of a stable,
nonrotating neutron star of mass $\SI{2}{\sunmass}$ \citep{Demorest2010,Antoniadis2013}.
Acronyms and references for the EOSs are listed in Table~\ref{tab:eos}, whereas
Figure~\ref{fig:eos} shows the $M = M(R)$ relation for each EOS. In addition, we
use a strange star EOS and two polytropic EOSs (for details, see below).

Ten of the thirteen EOSs are obtained by extracting
the zero-temperature limit from the tables provided by \citet{webhempel},
where they are available with full temperature  and composition dependence. 
We impose neutrino-less $\beta$-equilibrium, and therefore we choose the value of the lepton
fraction such that
\begin{equation}
  \label{eq:beta-equilibirum}
\mu_p + \mu_e = \mu_n + \mu_\nu \text{ with $\mu_\nu = 0$}\,,
\end{equation}
where $\mu_\nu,\mu_e,\mu_n,\mu_p$ are the chemical potentials for electron
neutrinos, electrons, neutrons and protons. The other three zero-temperature,
tabulated EOSs (WFF1, WFF2 and MDI) are provided in the distribution of the
public domain version of \texttt{RNS} \citep{rnscode}. Together, these thirteen
tabulated EOS cover the currently allowed mass-radius parameter space for
nonrotating models well, including very soft, very stiff and several
intermediate EOS, as well as an EOS\ with a large softening due to a phase
transition to hyperons, see Figure~\ref{fig:eos}.


The strange star EOS we consider is described by the MIT-bag model of quark
matter, with an EOS
\begin{equation}
  \label{eq:strange-eos}
  P(\epsilon) = a (\epsilon - \epsilon_0)\,,
\end{equation}
see, e.g.\ \cite{Witten1984,Haensel1986,Alcock1986,Zdunik2000}. The parameter
$\epsilon_0$, related to the bag constant, has a value of
$ \epsilon_0 = \SI{4.2785e14}{\g\per\cm\cubed}$. The value of $a$ is a
model-dependent constant which, assuming massless strange quarks, takes the
exact value $a = 1 \slash 3$. We fixed the surface of the star at the energy
density of $\SI{4.798e14}{\g\per\cm\cubed}$ (number density
$\SI{0.28665}{\per\femto\m}$). The $M(R)$ relation for nonrotating strange stars
with this EOS is substantially different than for the microphysically motivated
tabulated EOSs and it is marginally compatible with observational constraints.

Finally, we consider polytropic EOSs in the form
\begin{equation}
  \label{eq:polytropic-eos}
  P(\rho) = K \rho^{1 + \frac{1}{n}}\,,
\end{equation}
where $\rho$ is the rest mass density and $K$ and $n$ are respectively the
\emph{polytropic constant} and the \emph{polytropic index}. We included
polytropes with index $n = \num{0.5}$ and $n = \num{1.0}$, and we fixed the
value of $K = 1$, which corresponds to using the polytropic units as in
\citet{Cook1994}. The polytropic EOS with $n=\num{1.0}$ mimics roughly an
intermediate EOS, whereas the one with $n = \num{0.5}$ yields models that are
essentially a stiff core.

\newcolumntype{R}{>{\raggedleft\arraybackslash}X}%
\begin{table}
  \centering
  \caption{Equations of state with references used in this study.
    }
  \label{tab:eos}
  \begin{tabularx}{\columnwidth}{p{0.74cm}R}
    \textsc{EOS} & References                        \\     \hline
    DD2          & \citet{Hempel2010,Typel2012}      \\
    LS220        & \citet{Lattimer1991}              \\
    LS375        & \citet{Lattimer1991}              \\
    NL3          & \citet{Lalazissis1997,Hempel2010} \\
    SFHO         & \citet{Hempel2010, Steiner2013}   \\
    SFHX         & \citet{Hempel2010,Steiner2013}    \\
    TM1          & \citet{Sugahara1994, Hempel2010}  \\
    IUFSU        & \citet{RocaMaza2008, Hempel2010}  \\
    BHB$\Lambda$       & \citet{Hempel2010,Banik2014}  \\
    BHB$\Lambda \phi$ &   \citet{Hempel2010,Banik2014}   \\
    WFF1         & \citet{Wiringa1988}               \\
    WFF2         & \citet{Wiringa1988}               \\
    MDI          &  \cite{Arnett1977} \\
      \hline
  \end{tabularx}
\end{table}

\subsection{Equilibrium sequences and turning points}
\label{sec:equil-mod-turn}

We used \texttt{RNS} to construct a mesh of 500 different rotating equilibrium
models for each EOS, covering the range of central energy densities
$ \epsilon_c \in (\SI{0.5e15}{\g\per\cm\cubed}, \SI{5e15}{\g\per\cm\cubed})$
with every EOS in Table~\ref{tab:eos}. For polytropes, we used the range
$\epsilon_c \in (\num{0.3}, \num{3.0})$ for $n=\num{0.5}$ and
$\epsilon_c \in (\num{0.1}, \num{1.0})$ for $n=\num{1.0}$ (in polytropic units).
With \texttt{RNS} we could construct models that are either limited by the
mass-shedding limit, or that have a small axis ratio down to $\sim\num{0.3}$,
below which other branches of solutions with the same central energy density and
axis ratio may exist \citep[see][]{Rosinska2016, Ansorg2009}. Such models with
small axis ratio can have an off-center density maximum
\citep[][]{Stergioulas2004}. Because binary neutron star merger remnants are
quasi-spherical, with axis ratio that is not much smaller than about 0.4 in the
core , we did not include qusi-toroidal models in our sample. 

We interpolated with tenth-degree polynomials the obtained equilibrium sequences
in the region where we expected the turning points in the constant rest-mass and
constant angular momentum sequences. The turning points were then located by
imposing the stationary condition on the first derivative of the best-fitting
polynomials. We estimate an error of order $\num{0.5}\%$ in the central energy
density when locating the turning points.
\subsection{Validation of the numerical setup}
\label{sec:valid-numer-setup}

We verified our numerical setup by recovering the well-established result that
for uniformly rotating stars the location of the turning points of constant rest
mass sequences must coincide with the location of the turning points for
constant angular momentum sequences. We computed the turning points for both
kinds of sequences for all the EOSs in Table~\ref{tab:eos} and for the two
polytropic EOSs, with uniform rotation. Specifically, for each tabulated EOS we
constructed ten sequences with constant angular momentum with
$J \in (\num{0.5}, \num{7.0})$ and ten with constant rest mass (the range
depended on the EOS). For the polytropic EOS we computed twenty equispaced
sequences with parameters (in polytropic units)
$J \in (\num{0.5e-4}, \num{2.5e-2})$ and $ M \in (\num{0.15}, \num{0.19})$ for
the $n = \num{0.5}$ and $J \in (\num{0.5e-4}, \num{2.5e-2})$ and
$ M \in (\num{0.18}, \num{0.22})$ for $n = \num{1.0}$.

As expected,
we found that interpolating the turning points for the two kinds of sequences
with a fifth-order polynomial the two obtained curves are compatible within a
relative error of $0.5\%$ in the central energy density.
\section{Results}
\label{sec:results}

\subsection{Universal relations}
\label{sec:universal-relations}

\subsubsection{One-parameter rotation law}
\label{sec:one-param-rotat-1}

\begin{figure}
  \centering
  \setlength\figureheight{0.80\columnwidth}
  \setlength\figurewidth{\columnwidth}
  \includegraphics{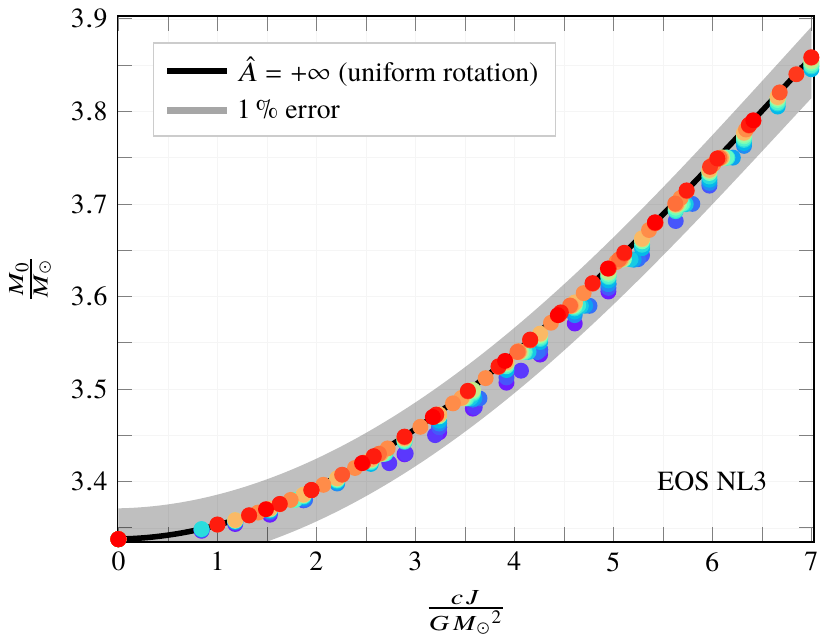}
  \caption{$\hat{A}$-independent relation between rest mass and angular momentum
    for the EOS NL3. This plot is obtained by finding the turning
    points for ten constant rest mass and angular momentum sequences with
    $\hat{A}$ between $\num{0.5}$ and $\num{2.0}$ and with values in the range
    $\{\num{2.5}, \num{3.0}, \num{4.0} \}$. Different colours represent
    different rotation parameters. Sequences with the same angular momentum (but
    different $\hat{A}$) divert vertically, whereas sequences with the same
    rest mass divert horizontal horizontally. The effect of decreasing $\hat{A}$ to lower the corresponding rest mass, and to increase the angular momentum. Overall this change is less than $\num{1}\%$
    , therefore all the sampled points lie on the same curve within this error bar. }
  \label{fig:universal-relation--J-M0-eosNL3.tikz}
\end{figure}

\begin{figure}
  \centering
  \setlength\figureheight{0.80\columnwidth}
  \setlength\figurewidth{\columnwidth}
  \includegraphics{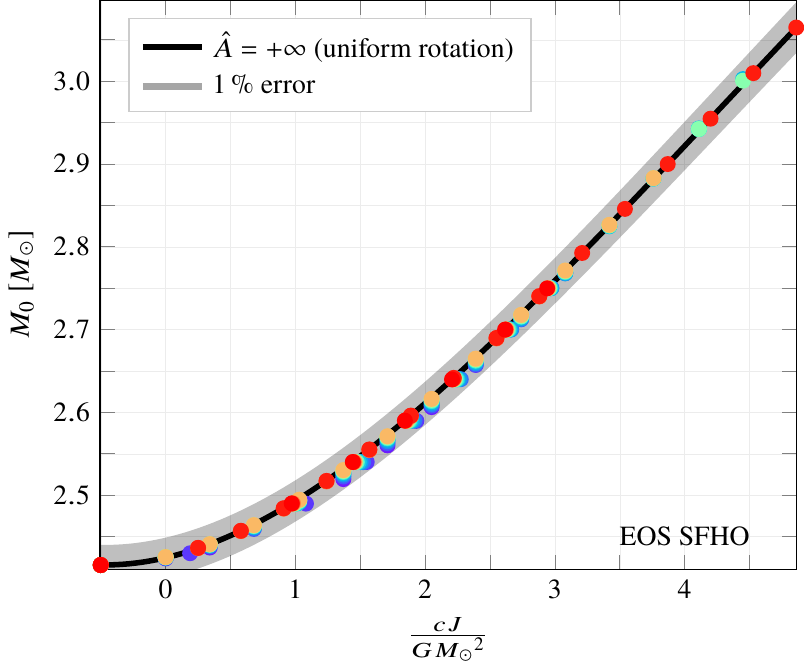}
  \caption{Same as Fig. \ref{fig:universal-relation--J-M0-eosNL3.tikz}, but for EOS SFHO.  }
  \label{fig:universal-relation--J-M0-eoSFHO.tikz}
\end{figure}

We studied the one-parameter rotation law with two representative cases of a soft
EOS (SFHO) and a stiff one (NL3). We sampled the parameter space
with ten equispaced values of $\hat{A}$ form $ \hat{A} = \num{0.5}$ to
$\hat{A} = \num{2.0} $ and with the values
$ \hat{A} \in \{\num{2.5}, \num{3.0}, \num{4.0} \}$. The locations of the
turning points with $\hat{A} = \num{4.0}$ are indistinguishable from the
uniformly rotating ones within the tolerance of our method. Therefore, we consider
that this value recovers uniform rotation with high accuracy and focus on a range of
$\hat{A}$ which produce results considerably different than  compared to uniform rotation.
We computed the turning points for ten equispaced equilibrium sequences of constant
angular momentum $J \in (\num{0.5}, \num{7.0})$ and for ten constant rest mass sequences with
$M_0 \in (\num{2.75}, \num{3.75})$ for NL3 and
$M_0 \in (\num{1.75}, \num{2.75})$ for SFHO. We found that there are relations
between their physical properties that depend only weakly on the parameter
$\hat{A}$. Examples of $\hat{A}$-independent relations that we found are
shown in Figures~\ref{fig:universal-relation--J-M0-eosNL3.tikz} and~\ref{fig:universal-relation--J-M0-eoSFHO.tikz}, where we find
 that for NL3 EOS and SFHO EOS the rest mass and the angular momentum are related
in a way which is practically independent of $\hat{A}$.

Since it is quite formidable to test every value of $\hat{A}$ for every single
one of the 13 different EOS in Table~\ref{tab:eos} (not only for the two we
tested), we conjecture that the $\hat{A}-$insensitive relations found for the
two representative EOS SFHO and NL3, will remain $\hat{A}-$insensitive also for
the other tabulated EOS. This means that \textit{universal (EOS-insensitive)
  relations valid for uniform rotation, should remain universal for differential
  rotation, if we can demonstrate that they are $\hat{A}-$insensitive for some
  representative EOS}.

For uniformly rotating models, we checked universality of relations for all EOS
by locating turning points of ten equispaced constant angular momentum sequences
with angular momentum from $\num{0.5}$ to $\num{7.0}$. We chose this range
because it covers all the allowed sequences for most EOS, as only the stiffest
ones can reach values of angular momentum greater than $\num{7.0}$ before
encountering the mass-shedding limit in uniform rotation, or before entering a
region in the parameter space where models have too small axis ratio to be
considered realistic for binary neutron star merger remnants. We computed only
constant angular momentum sequences, because the above range is suitable for all
the EOS, whereas the allowed range for rest masses depends on the stiffness of
the EOS so it would have required adjusting the range for each EOS.

First, for each EOS we found three relations involving the gravitational mass
$M$, the rest mass $M_0$, and the angular momentum $J$ that are
$\hat{A}-$insensitive with an error $\sigma_{\hat{A}}$ of
$\num{0.5} - \num{1.0}$ per cent. This means that if $x$ is a physical quantity
among the ones mentioned, and $f_{\hat{A}}$ one of the relations discovered,
then:
\begin{equation}
  \label{eq:fA_finf}
  \frac{\lvert f_{\mathrm{uni}}(x) - f_{\hat{A}}(x) \rvert} { \lvert f_{\mathrm{uni}}(x) \rvert } \leq
  \SI{1}{\percent} \quad \forall \hat{A} \geq \num{0.5}\,,
\end{equation}
where $f_{\mathrm{uni}}$ is the relation evaluated with uniform rotation
($\hat{A} = + \infty$).
The three   $\hat{A}-$insensitive relations we found for each EOS are simple  functions of the form:
\begin{subequations}
  \label{eq:A-equation}
  \begin{align}
   M_0 &= M_0(J),  \label{eq:A1} \\
   M_0 &= M_0(M), \label{eq:A2} \\
   M &= M(J).  \label{eq:A3}
  \end{align}
\end{subequations} For a fixed angular momentum $J$, the effect of decreasing
$\hat{A}$ is to lower the corresponding rest mass but to increase the
gravitational mass. On the other hand, for a fixed rest mass $M_0$, a reduction of
$\hat{A}$ produces an increase in the angular momentum and gravitational mass.
While the strength of the effect produced by varying $\hat{A}$ depends on the
stiffness of the EOS, it always stays smaller than $\num{1}\%$.


We obtain EOS-independent universal relations by rescaling relations~\eqref{eq:A1}, \eqref{eq:A2} and \eqref{eq:A3} using  $M^\star$ and $M_0^\star$ as follows:
\begin{subequations}
  \label{eq:universal-equation}
  \begin{align}
    \frac{M_0}{{M_0}^\star} &= 1 + 0.51
    \left(\frac{cJ}{G{{{M}^\star_0}}^2}\right)^2 - 0.28
    \left(\frac{cJ}{G{{{M}^\star_0}}^2}\right) ^4,  \label{eq:universal1} \\
    \frac{M_0}{{M}^\star_0} &= 0.93 \frac{M}{{M}^\star}
+ 0.07, \label{eq:universal2}\\
    \frac{M}{{M}^\star} &= 1
+ 0.29
    \left(\frac{cJ}{G{{M}^\star}^2}\right) ^2 - 0.10
    \left(\frac{cJ}{G{{M}^\star}^2}\right) ^4  \label{eq:universal3}.
  \end{align}
\end{subequations}

\begin{figure}
  \centering
  \setlength\figureheight{0.80\columnwidth}
  \setlength\figurewidth{\columnwidth}
   \includegraphics{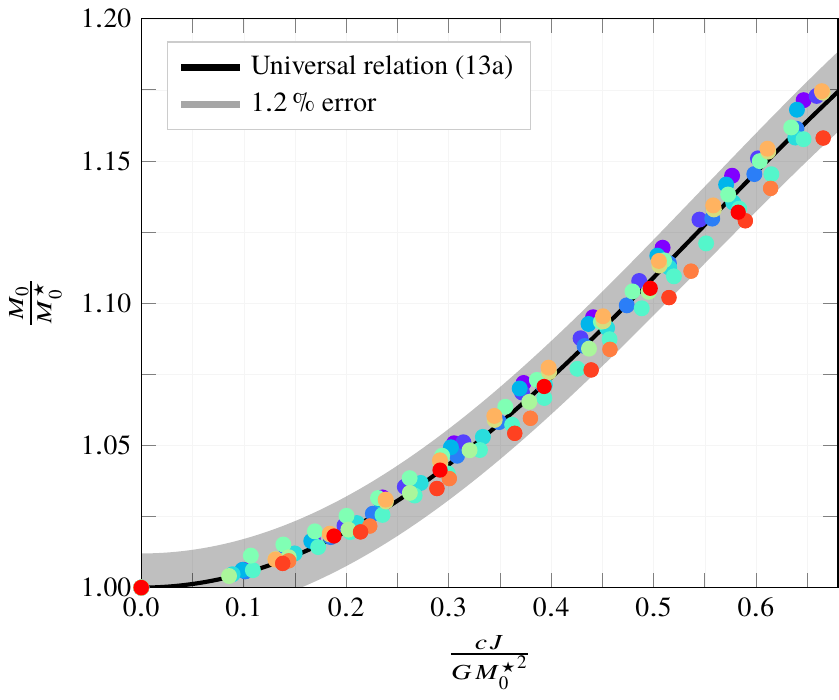}
   \caption{Universal (EOS-independent) relation between scaled rest mass and
     scaled angular momentum. The turning points for constant angular momentum
     sequences for uniformly rotating stars and different EOS, are shown with a
     different colour. All the points lie within a $\SI{1.2}{\percent}$ error bar.}
  \label{fig:universal-relation-J-M0}
\end{figure}

\begin{figure}
  \centering
  \setlength\figureheight{0.80\columnwidth}
  \setlength\figurewidth{\columnwidth}
   \includegraphics{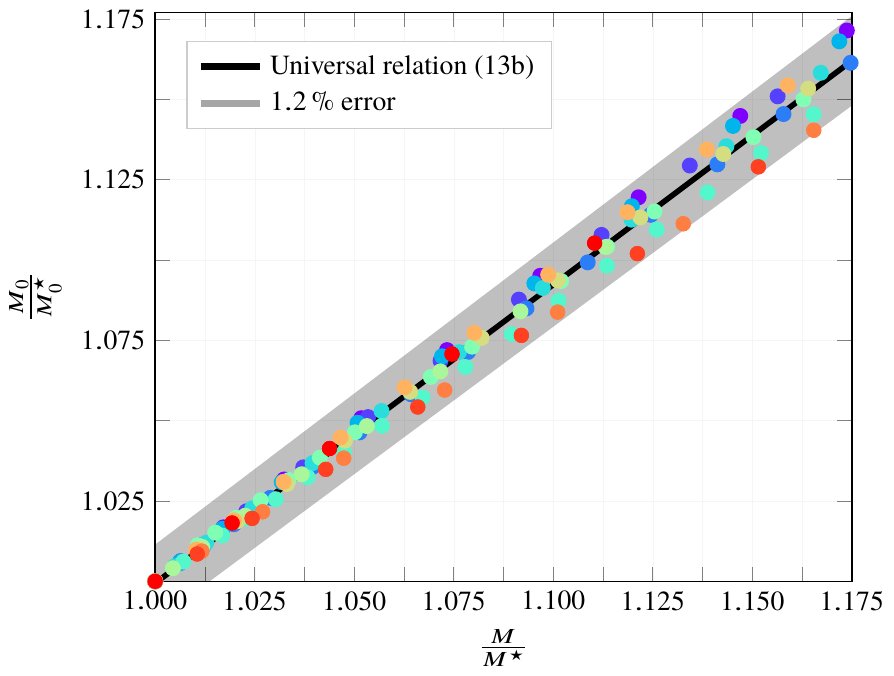}
   \caption{Universal (EOS-independent) relation between scaled rest mass and
     scaled gravitational mass. The turning points for constant angular momentum
     sequences for uniformly rotating stars and different EOS, are shown with a
     different colour. All the points lie within a $\SI{1.2}{\percent}$ error
     bar.}
  \label{fig:universal-relation-M-M0}
\end{figure}

\begin{figure}
  \centering
  \setlength\figureheight{0.80\columnwidth}
  \setlength\figurewidth{\columnwidth}
   \includegraphics{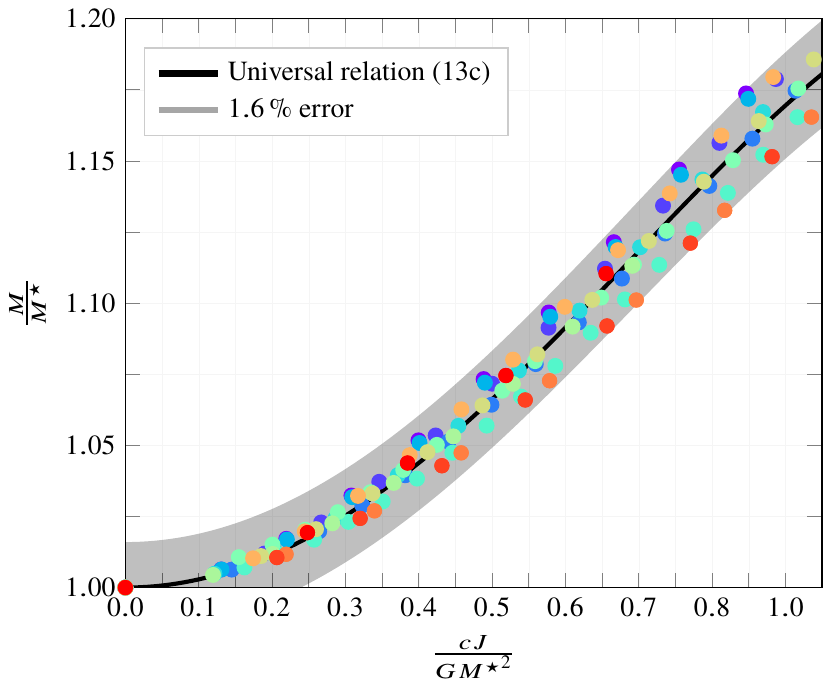}
   \caption{Universal (EOS-independent) relation between scaled gravitational mass and
     scaled angular momentum. The turning points for constant angular momentum
     sequences for uniformly rotating stars and different EOS, are shown with a
     different colour. All the points are within $\SI{1.6}{\percent}$ error
     bars.}
  \label{fig:universal-relation-J-M}
\end{figure}

As an example, we show (\ref{eq:universal3}) in
Figure~\ref{fig:universal-relation-J-M0} (black solid line). Every colour is a
different equation of state and the gray shade is the $\num{1.2}\%$ error bar
inside which every point lies. Similar maximum relative errors ($1.2\%$ and $1.6\%$) hold for the
other two universal relations \eqref{eq:universal1} and \eqref{eq:universal2}, which are shown in Figures \ref{fig:universal-relation-M-M0} and \ref{fig:universal-relation-J-M}.

Notice that, inverting relation \eqref{eq:universal1} we also find a universal
relation for the Kerr parameter $ a := c J \slash G M^2$ of the form:
\begin{equation}
  \label{eq:kerr-parameter}
  a = a \left( \frac{M}{M^\star} \right)
\end{equation}

Combining the maximum relative error of the universal relations for uniform
rotation with the maximum deviation due to differential rotation, we find that
relations \eqref{eq:universal1} and \eqref{eq:universal2} should be universal
for differentially rotating stars with a maximum relative error of about 2$\%$
(more specifically, for the range of astrophysically relevant values of $\hat A$
we considered with the one-parameter law (\ref{eq:one-parameter-law})).

\subsubsection{Three-parameter rotation law}
\label{sec:three-param-rotat-1}

In Section~\ref{sec:one-param-rotat-1} we used the one-parameter rotation law
(\ref{eq:one-parameter-law}) (and with uniform rotation as a special case). For
the three-parameter rotation law (\ref{eq:three-parameters-law}) we computed the
turning points for a finite subset of values of $\hat{A}_1, \hat{A}_2$ and
$\beta$. We combined in all the possible ways the values of
$\hat{A}_{1(2)} \in \{\num{0.5}, \num{1.0}, \num{2.5} \}$ and of
$\beta \in \{ \num{0.6}, \num{0.8} \},$ computing for each case eight sequences
of constant angular momentum $J \in (\num{0.5}, \num{7.0})$ and eight sequences
of constant rest mass $M_0 \in (\num{2.75}, \num{3.75})$ for the EOS NL3 and
$M_0 \in (\num{1.75}, \num{2.75})$ for the EOS SFHO.

In all cases studied, we confirmed that the universal relations presented in
Section~\ref{sec:one-param-rotat-1} still hold. This indicates that these
relations may retain universality for even more general rotation laws than
(\ref{eq:one-parameter-law}) or (\ref{eq:three-parameters-law}), at least for
quasi-spherical models considered here.


\subsubsection{Polytropic EOS}
\label{sec:polytr-equat-state}

We performed the same analysis of the Section~\ref{sec:one-param-rotat-1}
on the
two polytropic equations of state rotating with the one-parameter law finding
agreement with the previously presented relations. Moreover, polytropes showed
clearly how the stiffness of the equation of state affects the universal
relations.

\begin{figure}
  \centering
  \setlength\figureheight{0.80\columnwidth}
  \setlength\figurewidth{\columnwidth}
   \includegraphics{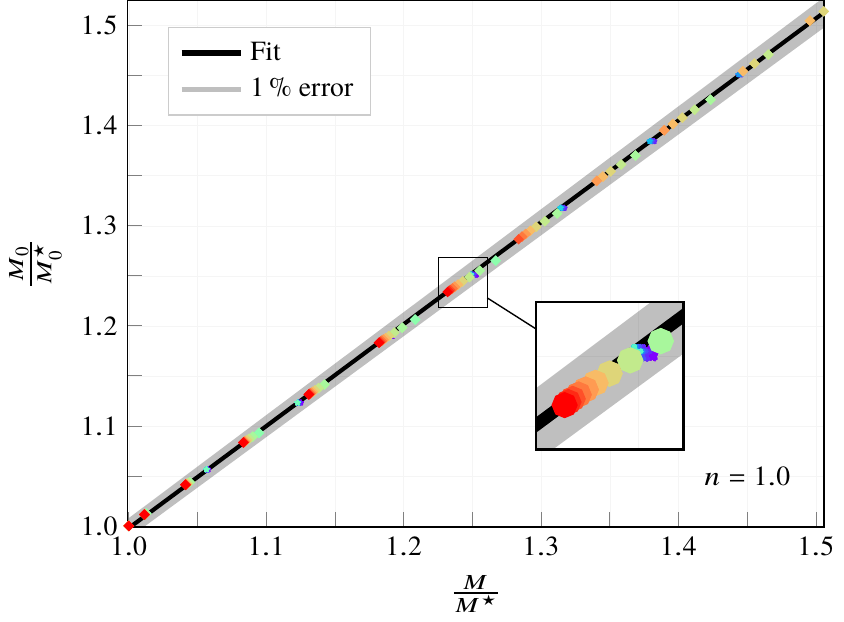}
  \caption{$\hat{A}$-independent relation between scaled  rest  mass and
scaled gravitational
    mass for a polytrope with index $n = \num{1.0}$. The relation is satisfied
    within a relative error of $\SI{1}{\percent}$. Nearby circles represent a sequence with the same angular momentum
    but different values of $\hat{A}$: decreasing $\hat{A}$ increases both
the
    rest mass and the gravitational mass in a way that the linear relation
is
    satisfied. The accuracy of this relation is $\num{0.3}\%$ , wheres it is $\num{1}\%$  in the case of the $n = \num{0.5}$ polytrope.}
  \label{fig:universal-other-eos-M-M0-p1-0}
\end{figure}

First, we noticed that for the $ n = \num{1.0} $ polytrope the relation between
rest mass and gravitational mass (depicted in
Figure~\ref{fig:universal-other-eos-M-M0-p1-0}) is especially accurate, with
error smaller than $\num{0.3}$ per cent, but somewhat larger for the
$ n = \num{0.5} $ polytrope. On the other hand, for the polytrope with
$ n = \num{0.5} $ the relation between gravitational mass and angular momentum
has accuracy of $\num{0.3}\%$, but it has $\sim\num{2}\%$ error in the case of
$ n = \num{1.0}$. In addition, the relation satisfied within
$\sim \SI{2}{\percent}$ shows clearly how the small dependence on $\hat{A}$
affects the physical quantities of the turning-point models (see inset of
Figure~\ref{fig:universal-other-eos-M-M0-p1-0}), which is similar to the effect
described in Section~\ref{sec:one-param-rotat-1} for tabulated EOS.

Having verified that there are some $\hat{A}$-independent
relations we checked whether polytropes satisfied
Equations~\eqref{eq:universal1},~\eqref{eq:universal2} and
\eqref{eq:universal3}. We found agreement, as shown  in
Figure~\ref{fig:universal-other-eos-J-M0}.

\subsubsection{Strange star EOS}
\label{sec:non-regul-equat}

\begin{figure}
  \centering
  \setlength\figureheight{0.80\columnwidth}
  \setlength\figurewidth{\columnwidth}
   \includegraphics{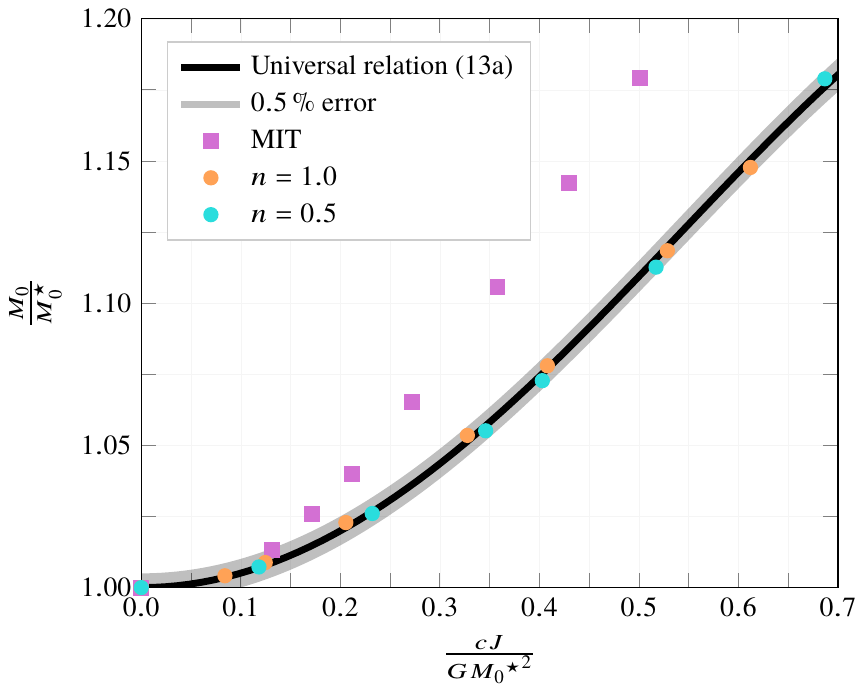}
  \caption{Scaled rest mass vs. scaled angular momentum relation for polytropic $n=0.5$ and $n=1.0$ and strange star (MIT) EOS. The polytropic EOSs satisfy the universal relation constructed with the tabulated EOSs with high accuracy. In contrast, the strange star EOS deviates significantly from the universal relation.}
  \label{fig:universal-other-eos-J-M0} 
\end{figure}

 Testing the relations with uniformly
rotating strange stars produced a counterexample: the turning points of constant
angular momentum sequences have a significantly different relation between rest
mass and angular momentum than for tabulated EOS (or polytropes). This situation is shown in
Figure~\ref{fig:universal-other-eos-J-M0}. A fit to the data obtained with the strange star EOS  is\begin{equation}
  \label{eq:strange-fit}
  \frac{M_0}{{M_0}^\star}\left(\frac{cJ}{G{{M}^\star_0}^2}\right) = 1 + 0.87
  \left(\frac{cJ}{G{{M}^\star_0}^2}\right) ^2 - 0.60
  \left(\frac{cJ}{G{{M}^\star_0}^2}\right) ^4\,.
\end{equation}
Equation~\eqref{eq:strange-fit} and Equation~\eqref{eq:universal1} are
compatible only in the limit of slow rotation, while for
$J\slash (M^\star_0)^2 = \num {0.9} $ the discrepancy is already $\num{15}\%$.

\section{SUMMARY}
\label{sec:conclusions}

Using the general-relativistic code \texttt{RNS} we investigated the properties of turning points of equilibrium sequences of differentially rotating
stars, using a variaty of EOS. Our interest is justified by the Friedman, Isper and Sorkin's method
that allows us to estimate the stability properties of rotating stars without
performing full dynamical simulations. 

Sampling the parameter space of the stellar models rotating with a
one-parameter law for differential rotation, we found that the properly scaled  gravitational mass, rest mass and angular
momentum of the turning points satisfy  universal relations that are independent of the EOS\ and the degree
of differential rotation to high accuracy.   A counterexample are strange stars, which do not obey the same universal relations as tabulated or polytropic EOS. We verified that the relations still hold for a more general, three-parameter rotation law, indicating 
that they might hold for an even larger class of rotation laws. 

Future work could extend our analysis to realistic rotation laws extracted
from dynamical simulations and the finite-temperature, pseudobarotropic models.

\section*{Acknowledgements}

Computations were performed on the \textsc{LCM} cluster by
INFN in Milan (Italy). We thank Matthias Hempel for providing EoS tables. A.B. acknowledges support by the Klaus Tschira Foundation.

\bibliographystyle{mnras}
\bibliography{abb}

\bsp    
\label{lastpage}
\end{document}